# Correlation-induced symmetry-broken states in large-angle twisted bilayer graphene on MoS$_2$

Kaihui Li[1], Long-Jing Yin[2], Chenglong Che[2], Xueying Liu[1], Yulong Xiao[1], Songlong Liu[2], Qingjun Tong[2,*], Si-Yu Li[1,3*], Anlian Pan[1,*]

[1]Key Laboratory for Micro-Nano Physics and Technology of Hunan Province, Hunan Institute of Optoelectronic Integration and College of Materials Science and Engineering, Hunan University, Changsha 410082, People's Republic of China

[2]School of Physics and Electronics, Hunan University, Changsha 410082, People's Republic of China

[3]Greater Bay Area Institute for Innovation, Hunan University, Guangzhou 511300, People's Republic of China

*Correspondence and requests for materials should be addressed to Q.J.T (e-mail: tongqj@hnu.edu.cn), S.Y.L (e-mail: lisiyu@hnu.edu.cn) and A.L.P. (e-mail: anlian.pan@hnu.edu.cn).

**Strongly correlated states are commonly emerged in twisted bilayer graphene (TBG) with "magic-angle" (1.1°), where the electron-electron (*e-e*) interaction *U* becomes prominent relative to the small bandwidth *W* of the nearly flat band. However, the stringent requirement of this magic angle makes the sample preparation and the further application facing great challenges. Here, using scanning tunneling microscopy (STM) and spectroscopy (STS), we demonstrate that the correlation-induced symmetry-broken states can also be achieved in a 3.45° TBG, via engineering this non-magic-angle TBG into regimes of *U*/*W* > 1. We enhance the *e-e* interaction through controlling the microscopic dielectric environment by using a MoS$_2$ substrate. Simultaneously, the bandwidth of the low-energy van Hove singularity (VHS) peak is reduced by enhancing the interlayer coupling via STM tip modulation. When partially filled, the VHS peak exhibits a giant splitting into two states flanked the Fermi level and shows a symmetry-broken LDOS distribution with a stripy charge order, which confirms the existence of strong correlation effect in our 3.45° TBG. Our result paves the way for the study and application of the correlation physics in TBGs with a wider range of twist angle.**

Twisted bilayer graphene (TBG) has attracted considerable interest as a paradigmatic 2D moiré material with two van Hove singularities (VHSs), which can be tuned by the twist angle $\theta$[1]. Near the "magic angle" (1.1°), the two VHSs merge into flat bands with almost vanishing Fermi velocity $v_F$[2]. Because of the dramatically reduced kinetic energy, the electron-electron (*e-e*) interaction becomes dominant in magic-angle TBG (MATBG), which results in many novel correlated states, such as superconductivity[3-5], nematicity[6] and magnetism[7-10]. However, these intriguing phenomena are mainly limited to samples with $\theta$ near the magic angle, which quickly diminish when $\theta$ deviates slightly due to the prompt increase of bandwidth $W$. The stringent requirement in fabrication process of MATBG with twist-angle precision below 0.1° limits greatly its generalization in fundamental physics and practical applications[11]. Actually, besides the twist angle, microscopic dielectric environment[12,13] and interlayer coupling[14-16] are also the two important factors that influence the *e-e* interaction $U$ and the bandwidth $W$. Engineering these two factors into the regimes of $U/W > 1$ provides another avenue to achieve strongly correlated states in the TBGs with $\theta$ away from the "magic angle".

In this letter, using scanning tunneling microscopy (STM) and spectroscopy (STS), we report the evidence of the strong correlation-induced symmetry-broken states in a 3.45° TBG on a $MoS_2$ substrate, through engineering the ratio $U/W$ to be about 1.6. On the one hand, the *e-e* interaction $U=e^2/4\pi\varepsilon L$ (where *e* is the electron charge) in this large-angle TBG is enhanced by the small permittivity $\varepsilon$ resulted from the dielectric $MoS_2$ substrate as well as its reduced moiré wavelength $L$. On the other hand, the bandwidth $W$ of the low-energy VHS peak is reduced by increasing the interlayer coupling using the STM tip modulation. When partially filled, the VHS LDOS peak in this 3.45° TBG exhibits a giant splitting, accompanied by the emergence of a symmetry-broken charge order. In the comparison studies of TBGs placed on conducting graphite substrate ($U/W < 1$), we observe no VHS splitting at various partial fillings, which confirms the decisive role of the correlation effect in resulting the VHS splitting in our 3.45° TBG.

Our STM measurements are performed on the TBG sample fabricated by the "tear and stack" method onto bulk $MoS_2$ at $T = 78$ K [Fig. 1(a), see Methods and Fig. S1 in Supplemental Material for details]. The bulk $MoS_2$ is chosen as the substrate of our TBG based on the following reasons. First, the semiconducting $MoS_2$ is a high-quality dielectric with a small permittivity $\varepsilon_{sub} = 4\varepsilon_0$[17] (where $\varepsilon_0$ is the vacuum permittivity) that can effectively reduce the screening effect (see Fig. S2 for the *dI/dV-V* spectra of the bulk MoS$_2$ substrate). The permittivity $\varepsilon$ of the TBG is determined by the average screening contribution of the dielectrics above and below it described as $\varepsilon = (\varepsilon_0 + \varepsilon_{sub})/2 = 2.5\varepsilon_0$[17,18]. Second, the large lattice mismatch between graphene and MoS$_2$ makes it impossible to form long-wavelength MoS$_2$-graphene moiré patterns that may influence the properties of TBG on top[12,19]. Third, the spin-orbit coupling in TBG induced by bulk MoS$_2$ via proximity effect is much less than 1 meV, which has little influence on the electronic property of the TBG[20,21]. Fig. 1(b) gives the STM topography detected on the TBG sample, which shows the moiré superlattices with the period $L \sim 4.1$ nm (see Fig. S3 for more STM images). The interlayer twist angle is estimated as $\theta \sim 3.45°$ according to the formula $L=a/[2\sin(\theta/2)]$ with the graphene lattice constant $a = 0.246$ nm[22,23]. Due to the decreased permittivity $\varepsilon$ and the small moiré wavelength $L$, the effective Coulomb interaction $U$ of our 3.45° TBG is dramatically enhanced to be about 146 meV. We have adopted the formula $U=e^2/4\pi\varepsilon L$ to estimate the *e-e* interaction of our 3.45° TBG, which is widely used in MATBG and other related moiré systems[12,19,24-27], because of the similarity in localization behavior of their electronic states (see Fig. S4 for details).

Figure 1(c) shows a typical *dI/dV-V* spectrum of the 3.45° TBG, exhibiting two low-energy VHSs (VHS$_1$ and VHS$_2$) and two remote bands flanked the Dirac point (see Fig. S5 for more spectra). We can make two direct observations from the spectra. First, the energy separation $\Delta E_{VHS}$ of the two VHSs is measured to be about 254 meV. Generally, $\Delta E_{VHS}$ is determined by both the twist angle $\theta$ and the interlayer coupling $t_\theta$ described as $\Delta E_{VHS} \approx \hbar v_F \Delta K - 2t_\theta$, where $\hbar$ is the reduced Planck constant, $\Delta K=2|K|\sin(\theta/2)$ is the momentum separation between the two Dirac cones of the two graphene monolayers and $|K|=4\pi/3a$[22,23,28] [see the inset of Fig. 1(a)]. Consequently, we estimate the interlayer coupling of our 3.45° TBG as $t_\theta \sim 212$ meV. Second, the bandwidth $W$, defined as the full width at the half maximum (FWHM) of the VHS LDOS peak, is measured to be $111.8 \pm 3.1$ meV for VHS$_1$ and $122.0 \pm 2.8$ meV for VHS$_2$. These two observations are consistent with that previously reported in TBGs with $\theta \sim 3°$ which are treated to be in a weakly correlated regime and described by single-particle model[23,28]. With the measured $t_\theta \sim 212$ meV, we calculate the band structure and LDOS of 3.45° TBG using the single-particle model in Figs. 1(d), 1(e) (see Method in Supplemental Material). The calculated energy location and FWHM of the two VHSs and two remote bands well reproduce our experimental results in Fig. 1(c). We estimate the ratio $U/W$ of our 3.45° TBG to be about 1.3 for VHS$_1$ and 1.2 for VHS$_2$, which are slightly larger than 1.

According to previous studies, both the energy separation $\Delta E_{VHS}$ of the two VHSs and the bandwidth $W$ of their LDOS peaks would decrease with the increase of interlayer coupling $t_\theta$ in large-angle TBGs[14-16]. Furthermore, the $t_\theta$ depends exponentially on the interlayer spacing $d$ described as $t_\theta \propto e^{-C(d-d_{eq})}$, where $C$ is a constant and $d_{eq} \sim 0.34$ nm is the equilibrium spacing[28,29]. Therefore, the bandwidth $W$ can be experimentally reduced by shortening the interlayer spacing of TBG. Arising from the atomic-thick nature of graphene systems, the STM tip modulation method has been widely used in manipulating their atomic structure and physical property[29-34]. Fig. S6 in Supplemental Material and Fig. 1(f) compare the STM topographies and the background-subtracted $dI/dV$-$V$ spectra of the 3.45° TBG taken before and after a STM tip pulses (3V, 100 ms), respectively. In the STM images under the same measurement condition (Fig. S6), the height difference between the AA and AB/BA stackings increases about 20 pm after the tip pulse, resulted from the increased interlayer potential [30,35]. The FWHMs (bandwidth $W$) of the VHSs after the tip pulse are reduced to 91.8 ± 3.2 meV for $VHS_1$ and 90.3 ± 2.5 meV for $VHS_2$. And the VHSs separation $\Delta E_{VHS}$ decreases to 233.1 ± 3.2 meV, therefore, the $t_\theta$ is enhanced to 223 meV after the tip modulation [see Fig. S7 for more spectra after the tip pulse]. Overall, the ratio $U/W$ increases to about 1.6 in our 3.45° TBG after the above optimization of the dielectric environment and the enhancement of the interlayer coupling. This value is comparable to the ratio $U/W = 1.4 - 3$ estimated in the MATBG[36-39], which makes it possible to observe prominent correlation effect in our 3.45° TBG.

To study the strongly correlated states in our 3.45° TBG, the VHS LDOS peak needs to become partially filled. This is achieved because the inhomogeneity of the local charge and the mechanical deformation in the underlying $MoS_2$ substrate will introduce a doping variation in the TBG sample through the charge transfer[40,41] (see Fig. S3 for details of doping variation). Taking this doping inhomogeneity as our advantage, we can investigate the surface state at different electron fillings by simply moving the STM tip to different locations of the sample. We note that the topography of the moiré superlattice shows little deviation at different regions (Fig. S3), which excludes the possible influence of local strain. Fig. 2(a) shows the $dI/dV$-$V$ spectra of our 3.45° TBG taken from different locations, where the LDOS peaks shift with different doping levels (see Fig. S8 for more spectra). Interesting, as soon as the Fermi level (zero bias) enters the $VHS_2$ LDOS peak, this peak broadens and starts to split centered on the Fermi level. We note that this VHS splitting is not resulted from the Altshuler–Aronov zero-bias anomaly[42], because there is neither nanoscale impurity nor disorder in our 3.45° TBG [Fig. 1(b) and Fig. S3].

In order to better analyze the split behavior at different fillings, we plot the $VHS_2$ LDOS peaks within a smaller energy range after subtracting the background from the *dI/dV-V* spectra in Fig. 2(b) (see Fig. S9 for the original spectra and the background subtraction process). Through performing the Gaussian fittings, we observe that the deformed $VHS_2$ peaks at partial fillings split into two peaks flanked the Fermi level, which gives rise to a pseudo-gap-like feature. These two split peaks are labeled as lower band (LB) and upper band (UB), the relative height of which varies with different doping levels. A similar correlation-induced splitting with a pseudo-gap of about 10 meV was previously observed in partially filled flat-band of MATBG[38,39,43,44], which indicates that the correlation effect also plays a prominent role in our 3.45° TBG. We estimate the local charge density by the spectral weight of the two split peaks below and above the Fermi level, which has been conventionally adopted in the study of flat band systems[41,43]. The filling ratio is defined as the ratio of the area $A_{LB}$ under the *dI/dV-V* spectrum spanned by the LB to the total area $A_{TOT}$ spanned by the whole $VHS_2$[41]. Fig. 2(c) exhibits a typical example of a half-filled $VHS_2$ peak which splits into two peaks with almost the same coverage area (see Fig. S10 for the $VHS_2$ spectra with other filling ratios). As shown in the filling-ratio-dependent STS map [Fig. 2(d)], the $VHS_2$ splits into two peaks once the filling ratio is larger than zero, which provides direct evidence for the existence of strong correlation effect in our 3.45° TBG. The splitting magnitude $\Delta E_S$ of the partially filled $VHS_2$ is summarized to be around 76 meV and shows little dependence on the filling ratio in Fig. 2(e). The $\Delta E_S$ here is much larger than the splitting of partially filled flat-band in MATBG because of the large Coulomb interaction $U$ in our 3.45° TBG. We note that, before the tip pulse modulation, the $VHS_2$ LDOS peaks at partial fillings exhibit a less-obvious splitting tendency because of the relatively small $U/W$ ratio (see Fig. S11 for details).

To highlight the importance of microscopic dielectric environment in realizing the above correlated states, we perform the comparison experiments on the TBGs placed on highly oriented pyrolytic graphite (HOPG) substrate (Fig. 3). The conducting graphite substrate with a large permittivity $\varepsilon_{HOPG} \sim 15$ can strongly screen the *e-e* interaction in the TBGs placed on top[45,46]. Fig. 3(a) shows the STM image of a 1.93° TBG on HOPG substrate with the moiré period of 7.3 nm. Its *dI/dV-V* spectra in Fig. 3(b) shows that the two VHSs shift with the varying of doping level (see Fig. S12 for more spectra). We summarize the FWHM of the $VHS_2$ ($VHS_1$) peak as a function of the filling ratio (the energy position) in Fig. 3(c) (Fig. S12). Different from the case of TBG on $MoS_2$, we observe neither FWHM broadening nor peak splitting for the partially filled $VHS_2$ and the fully filled $VHS_1$. Similar result is also observed in the 1.88° TBG on HOPG as shown in Fig. S13, where the Fermi level enters the $VHS_1$ peak. Considering the peak FWHM and the HOPG substrate permittivity, we estimate the $U/W$ in the 1.93° and 1.88° TBGs are around 0.26 and 0.33 respectively, which indicates that the correlation effect is weak in these two samples. Therefore, it is reasonable for the absence of strong-correlation-induced VHS splitting in these two samples. We also perform the tip pulse modulation in the 1.93° TBG on HOPG to enhance the interlayer coupling (see Fig. S14 for details), however, no VHS splitting is observed, which in turn reflects the dominate role played by screening effect from the conducting substrate in suppressing the correlated states in this case.

The above results demonstrate that it is possible to introduce strong correlation effect in large-angle TBG by engineering the ratio $U/W > 1$. We now turn to explore the resulting novel quantum states via studying the spatial LDOS distribution in the 3.45° TBG on bulk $MoS_2$ using STS mapping. Fig. 4(a) shows the STS mapping obtained at 83 meV corresponding to the energy of the empty $VHS_2$, where the LDOS distribution reveals the same period as the moiré superlattices in Fig. 1(b). The moiré-induced LDOS distribution can also be observed in the Fast-Fourier-transform (FFT) image of the STS mapping [see inset of Fig. 4(a)]. The relative magnitude of the differential conductance of the moiré-induced LDOS distribution in Fig. 4(a) is not as strong as that in MATBG, because of the weaker electron localization in our 3.45° TBG. To clearly image the shape and symmetry of the LDOS distribution, we do the inverse FFT of the filtered inner hexagonal spots in the FFT image of Fig. 4(a)[47,48], and the corresponding FFT-filtered image is given in Fig. 4(b). We observe the LDOS distribution exhibits a nearly rotational symmetry that is the same as the spatial moiré superlattices. The STS mapping of the empty $VHS_2$ in Figs. 4(a), (b) can also help to rule out the possible influence of tip anisotropy and strain effect.

For the case of the half-filled $VHS_2$, the LDOS distribution is shown in the FFT-filtered images of Figs. 4(c), (d), which are extracted from the STS mappings taken at the energies of the two split peaks of $VHS_2$ (see Fig. S15 for the corresponding original STS mappings). Interestingly, we observe a globally stripy charge order in the LDOS distribution images, which aligns with the crystallographic axis of the moiré superlattices. This stripy charge order breaks the rotational symmetry of the moiré superlattice, indicating that the *e-e* interaction generates a symmetry-broken phase when the VHS is partially filled in our 3.45° TBG. This interesting symmetry-broken phase, as conventionally observed in MATBG[38,43,49], is the first time to be spatially detected in a large-angle TBG.

In summary, through controlling the microscopic dielectric environment by the $MoS_2$ substrate and reducing the bandwidth by the STM tip modulation, we realize a strongly correlated state in a 3.45° TBG which was previously believed to be in a weakly correlated regime. Our results demonstrate an effective way to introduce correlated states in large-angle TBGs, which opens up a large "twist-angle window" in TBG system to study and apply the correlation physics besides of the famous "magic angle".

**Acknowledgements:**

This work was supported by the National Key R&D Program of China (Nos. 2022YFA1204700, 2022YFA1204300); the National Natural Science Foundation of China (Nos. 12104144, 52221001, 62090035, U19A2090); the National Natural Science Foundation of Hunan Province, China (Grant Nos. 2021JJ20025, 2022JJ10002); the National Natural Science Foundation of Guangdong Province, China (No. 2023A1515012699); the Key Program of Science and Technology Department of Hunan Province (Nos. 2019XK2001, 2020XK2001).

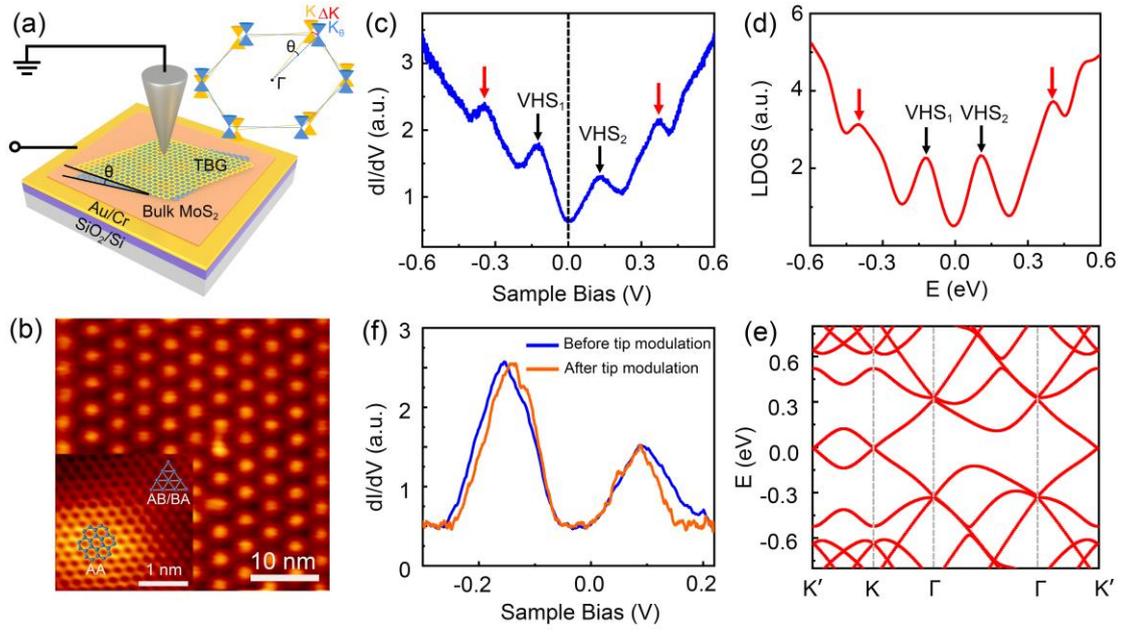

**FIG.1 (a)** Schematic of STM measurement setup on TBG supported by bulk MoS$_2$ substrate. Inset: schematic of two Dirac cones of the two graphene layers intersecting in momentum space. **(b)** A 40nm×40nm STM image of the 3.45° TBG ($V_{Sample}$ = 180 mV and $I$ = 100 pA). Inset: An atomic-resolved STM image of the moiré superlattice. **(c)** A typical *dI/dV-V* spectrum acquired at the AA-stacked region. The black and red arrows mark the two VHSs and two remote bands. **(d), (e)** Theoretical LDOS and band structure of 3.45° TBG. **(f)** The background-subtracted *dI/dV-V* spectra taken before and after a STM tip pulse, respectively.

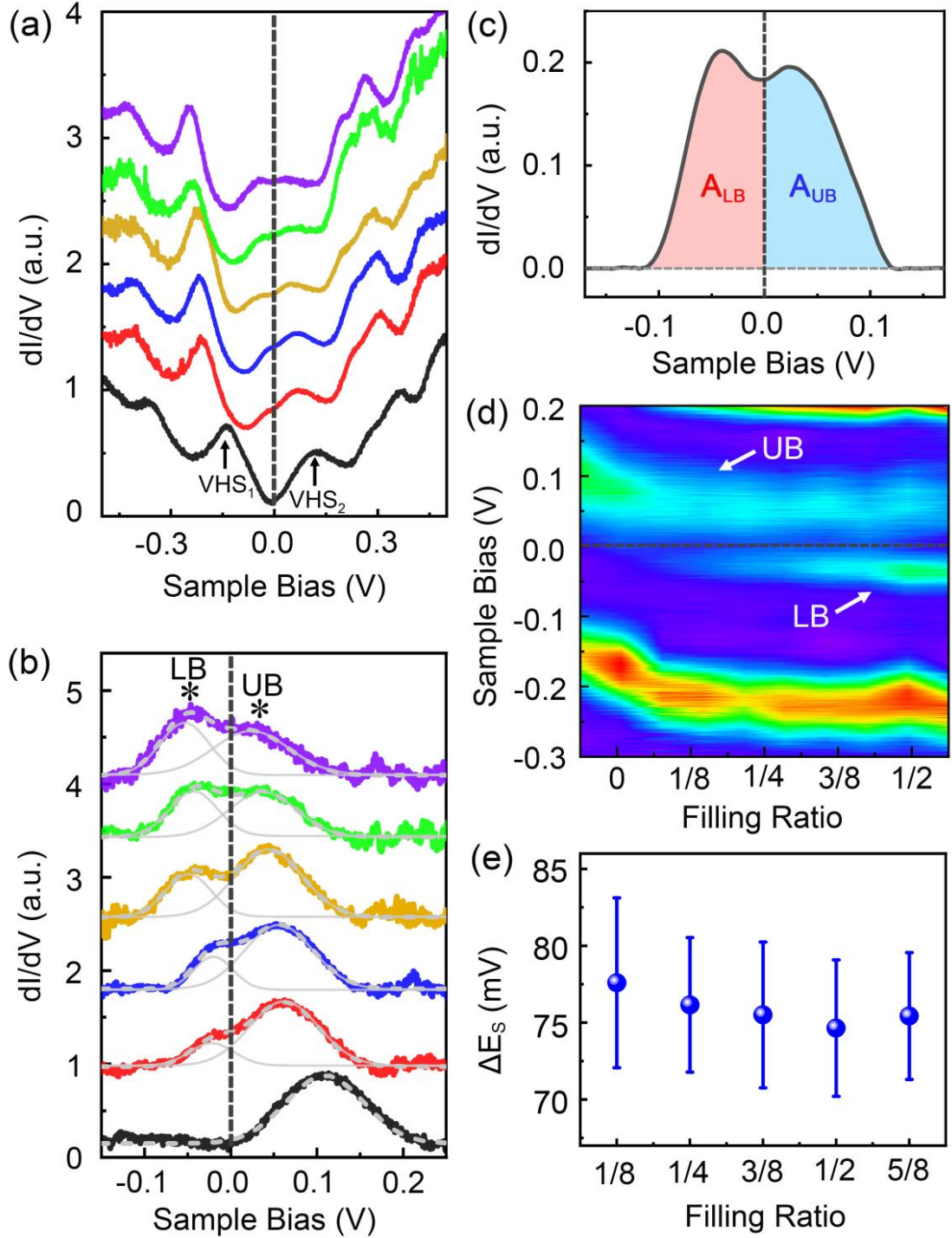

**FIG.2 (a)** Representative $dI/dV$-$V$ spectra of the 3.45° TBG with different doping levels, measured at different locations of the sample. The spectra are offset in the $Y$ axis for clarity. The black dashed line marks the position of the Fermi level. **(b)** The evolution of the $VHS_2$ LDOS peaks within a smaller energy range after subtracting the background from the $dI/dV$-$V$ spectra. The grey lines are the Gaussian fittings. **(c)** A background-subtracted $dI/dV$-$V$ spectrum of the half-filled $VHS_2$. Colored areas $A_{LB}$ and $A_{UB}$ mark the ones spanned by the LB and UB, respectively. **(d)** The color map of filling-ratio-dependent *dI/dV-V* spectra. **(e)** Summarized splitting magnitude $\Delta E_S$ of the partially filled VHS$_2$ as a function of the filling ratio.

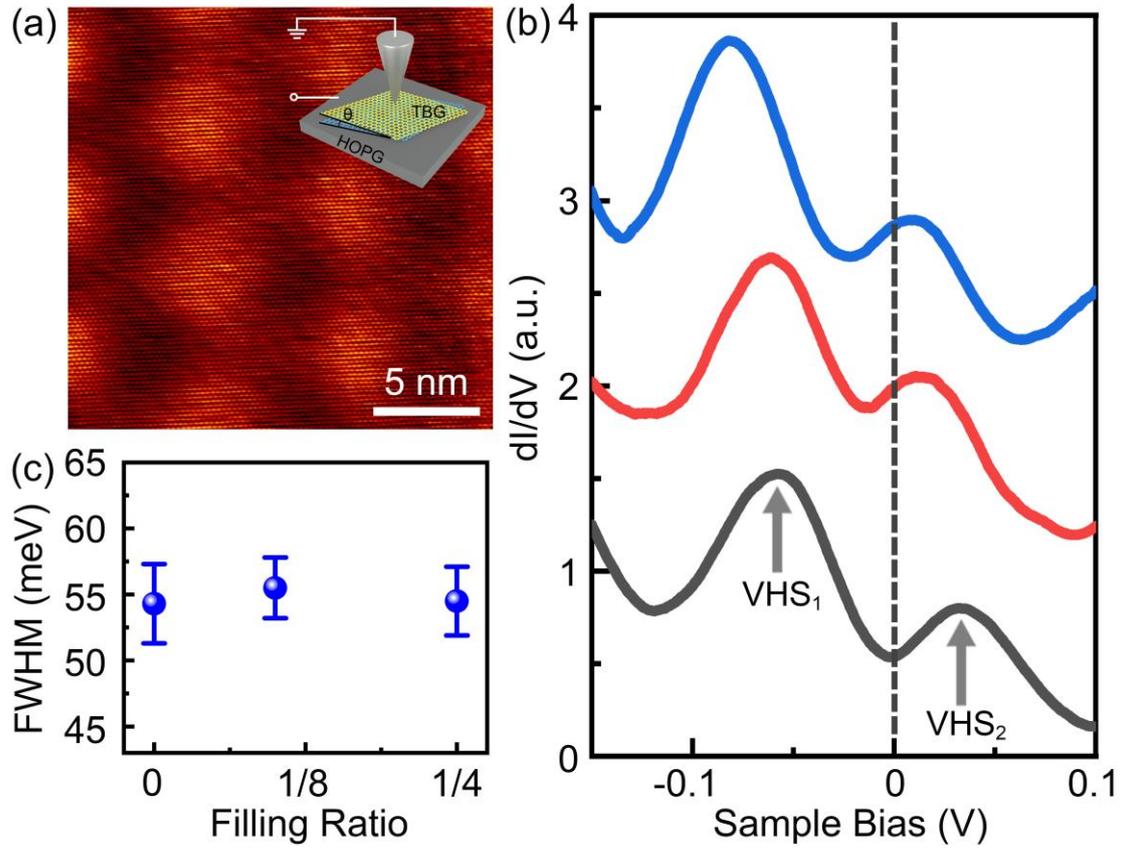

**FIG.3 (a)** A 20 nm×20 nm STM image of a 1.93° TBG on a HOPG substrate. Inset: Schematic of STM measurement setup. **(b)** The *dI/dV-V* spectra obtained at AA stacked region with different doping levels. The black dashed line marks the position of the Fermi level. **(d)** Summarized FWHM of the VHS$_2$ LDOS peaks as a function of the filling ratio.

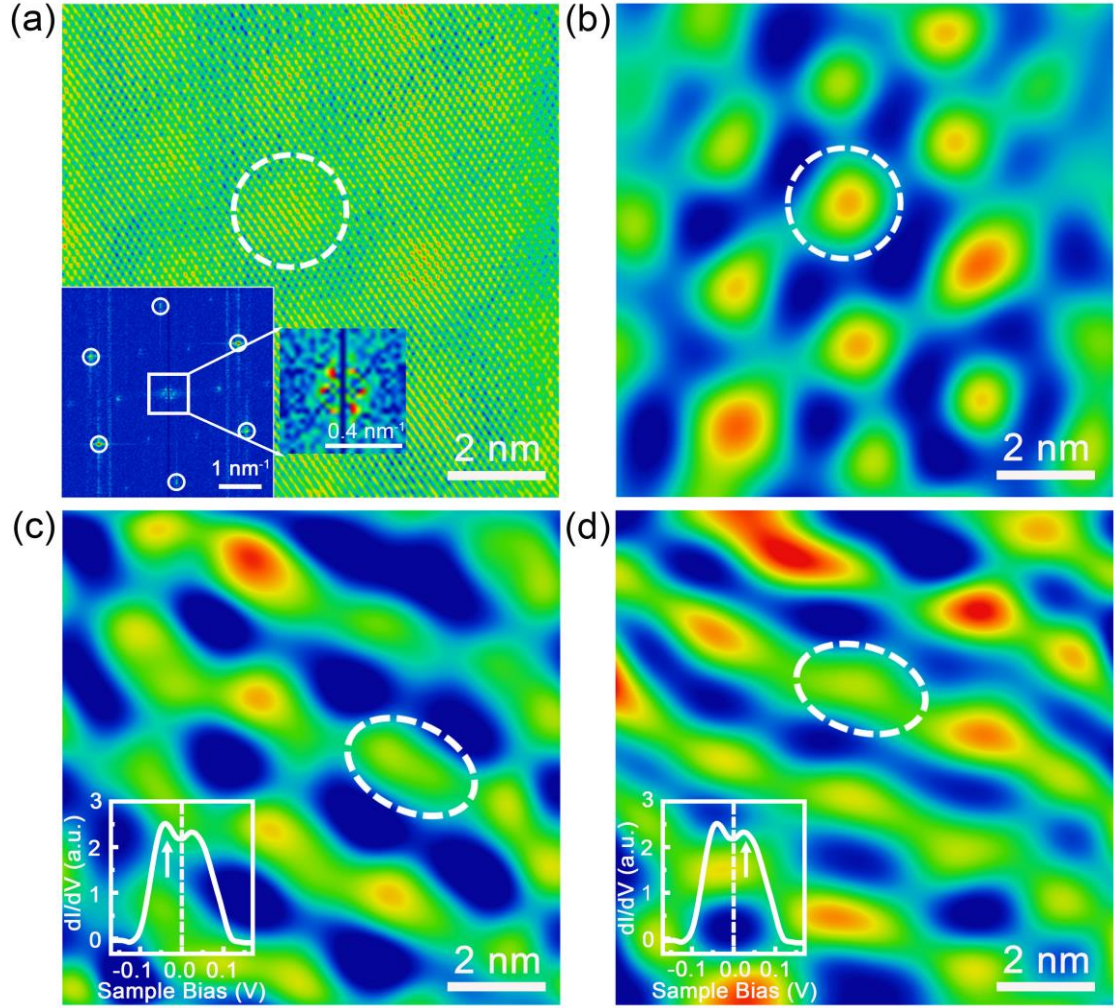

**FIG.4 (a)** A 15 nm×15 nm STS mapping obtained at 83 meV corresponding to the energy of the empty VHS$_2$ in the 3.45° TBG. Inset: The corresponding FFT image. The outer and inner hexagonal spots correspond to the graphene lattice and the moiré-induced LDOS distribution, respectively. **(b)** The FFT-filtered image of the STS mapping in panel (a) **(c), (d)** The 15 nm×15 nm FFT-filtered images extracted from the STS mappings taken at the energies of the two split peaks of the half-filled VHS$_2$ (-40 meV and 40 meV), respectively. The white loops in the STS mappings mark the patterns of the LDOS distribution.